\documentclass[preprint,12pt]{aastex}

\usepackage{epsfig}
\usepackage{graphicx}

\def\gx{GX~339$-$4}

\def\grs{GRS~1915$+$105}
\def\gro{GRO~J1655$-$40}

\def\1e{1E~1740.7$-$2942}

\def\xte{XTE~J1550$-$564}
\def\xt{XTE~J1650$-$500}
\def\xteonze{XTE~J1118$+$480}
\def\1h{H~1743$-$322}
\def\4u{4U~1755$-$33}

\slugcomment{ApJ, in press.}

\shorttitle{Black holes in quiescence.}
\shortauthors{Corbel et al.}

\begin{document}

  \title{On the origin of black hole X-ray emission in quiescence: {\em Chandra} observations of  \xte\ and \1h .}

   \author{S. Corbel\altaffilmark{1}, J.A. Tomsick\altaffilmark{2}, P. Kaaret\altaffilmark{3}}
   \altaffiltext{1}{AIM - Unit\'e Mixte de Recherche CEA - CNRS - Universit\'e Paris VII
        - UMR 7158, CEA Saclay, Service d'Astrophysique, F-91191 Gif sur Yvette, France.}
   \altaffiltext{2}{Center for Astrophysics and Space Sciences,
                University of California  at San Diego, MS 0424, La Jolla, CA92093, USA.}
   \altaffiltext{3}{Department of Physics and Astronomy, University of Iowa, Iowa City, IA 52242 USA.}

\begin{abstract}

	We report the results of observations of the black hole binaries \xte\ and \1h\ in their 
	quiescent state using the {\em Chandra X-ray Observatory}. Both sources are detected at their 
	faintest level of X-ray emission ever observed with a 0.5--10 keV unabsorbed luminosity of 2 $\times$ 
	10$^{32}$ (d/5 kpc)$^2$ erg s$^{-1}$ for \xte\ and 9 $\times$ 10$^{31}$ (d/8 kpc)$^2$ erg s$^{-1}$ for \1h. 
	These luminosities are in the upper range compared to the faintest levels observed in other black hole 
	systems, possibly related to residual accretion for these sources with frequent outbursts. For \xte, 
	the {\em Chandra} observations also 
	constrain the X-ray spectrum as a fit with an absorbed power-law model yields a photon
	index of 2.25 $\pm$ 0.08, clearly indicating a softening of the X-ray spectrum at lower 
	luminosities compared to the standard hard state. Similar softening at low luminosity is seen
	for several black hole transients with orbital periods less than 60 hours. Most of the current 
	models of accreting black holes are able to reproduce such softening in quiescence.
	In contrast, we find that systems with orbital periods longer than 60 hours appear to have 
	hard spectra in quiescence and their behaviour may be consistent with hardening in quiescence. 

\end{abstract}


\section{Introduction}

X-ray Novae (or Soft X-ray Transients) are compact binaries in which a neutron star
or black hole (BH) primary accretes from a donor star via Roche-lobe over-flow.  Most 
of these systems are usually in a quiescent state with an X-ray luminosity of 10$^{30}$ 
to 10$^{33}$ erg s$^{-1}$. However, they undergo episodic outbursts that last for 
months with X-ray luminosities that can sometime 
reach or exceed the Eddington limit ($\sim$ 10$^{39}$ erg s$^{-1}$ for a 10 M$_\odot$ BH).
Despite the residual activity of these quiescent objects, very little is known about
their emission properties at very low accretion rates (McClintock \& Remillard 2005). 
With the sensitivity of current X-ray missions (in particular {\em Chandra} and 
{\em XMM-Newton}), it is now possible to study in more detail the physical processes 
that take place in this accretion regime.

\xte\ was discovered by {\em RXTE}/ASM on 1998 September 7 (Smith 1998). 
A brief and intense radio/X-ray flare, associated with a massive plasma ejection,
was observed two weeks later (Hannikainen et al. 2001). Subsequent radio and
X-ray observations revealed the formation of large scale jets moving away
from the \xte\ black hole over the course of several years (Corbel et al. 
2002; Tomsick et al. 2003a; Kaaret et al. 2003). After its discovery outburst 
in 1998--1999, \xte\ had a second strong outburst in 2000, and fainter and 
shorter outbursts were detected in 2001, 2002 and 2003 (Figure 1).  Optical 
observations indicate that the compact object in \xte\ is likely a black hole
of 10.5 $\pm$ 1.0 M$_\odot$ at a distance of about 5.3 kpc (Orosz et al. 2002).

\1h\ was discovered with {\em Ariel V} in August 1977 (Kaluzienski \& Holt 1977) 
and was precisely localized by {\em HEAO-1} a few weeks later (Doxsey et al. 1977). 
Based on its X-ray properties, \1h\ has been classified as a black hole candidate 
(White \& Marshall 1983).  In March 2003, {\em INTEGRAL} detected new activity 
from IGR~J17464$-$3213 (Revnivtsev et al. 2003) that was later found to correspond 
to \1h.  During outburst, \1h\ went through several X-ray states with properties 
typical of BHC. The 2003 outburst ended around late November 2003. \1h\ was observed 
again in outburst from July 2004 (Swank 2004) to November 2004. A bright radio flare 
(likely associated with a massive ejection event) was observed in 2003 by Rupen et 
al. (2003). The ejected plasma was later found to interact with the interstellar 
medium causing in-situ particle acceleration and the formation of two large-scale, 
synchrotron-emitting radio and X-ray jets (Corbel et al. 2005) as in the \xte\ case. 

In this paper, we present the results of ten {\em Chandra} observations of the
two black holes \xte\ and \1h\ carried out during their quiescent state. These two
sources have been detected at their faintest level ever observed. The high quality
{\em Chandra} spectra for \xte\ allow us to study  and monitor the emission 
properties of this black hole at very low accretion rate. We then investigate
the quiescent emission of black hole systems. 


\section{Observations and data reduction}

Our {\em Chandra} program of monitoring X-ray jets from \xte\ and \1h\ has also 
allowed us to study in great detail the black holes in these two systems. \xte\ 
was observed by {\em Chandra} on five occasions: 2002 March 11 (MJD 52344.8), 2002 June 19 
(MJD 52444.5), 2002 September 24 (MJD 52542.0), 2003 January 27 (MJD 52667.3) 
and 2003 October 23 (MJD 52935.6).  For completeness, we also include the results 
of two published {\em Chandra} observations on 2000 August 21 (MJD 51777.4) and 
2000 September 11 (MJD 51798.3) that were performed during the decay of the 2000 
outburst of \xte\ (Tomsick et al. 2001). The {\em Chandra} observations of \1h\ 
were carried out on three occasions: 2004 February 12 (MJD 53048), March 24 (MJD 53089) and 
March 27 (MJD 53092). Figure~1 shows the {\em RXTE}/ASM
1.5--12 keV light-curves of \xte\ and \1h, and the arrows indicate the dates of
the {\em Chandra} observations.  This illustrates that all our observations were
conducted a few months (or even up to a year for \xte) after or before  a period
of significant X-ray activity (outburst).
This allows us to monitor the emission properties of \xte\ and \1h\ during their
quiescent phases.

For all {\em Chandra} observations, we used the Advanced CCD Imaging Spectrometer 
spectroscopic array (ACIS-S; Bautz et al. 1998) in imaging mode, with the target 
placed on one of the back-illuminated ACIS Chips (S3).  For the first observation of 
\xte\ in 2002 and the first observation of \1h, only the S3 chip was read 
out and a 1/2 sub-array mode was used to 
limit pile-up.  For the later observations, the sources were known to be at lower 
flux and the full ACIS-S imaging mode array was used.

We produced 0.3--8 keV ACIS images using the ``level 2'' event lists from the 
standard data processing using the {\em Chandra} Interactive Analysis of 
Observations (CIAO) software package.  We constructed light curves with all 
valid events on the S3 chips to search for times of high background. Periods 
with background flares were removed using the standard CIAO script {\em analyse\_ltcrv.sl}  . 
We searched for X-ray sources in each 
0.3--8 keV image using {\it wavdetect} (Freeman et al.\ 2002), the wavelet-based 
source detection routine in CIAO. For all {\em Chandra} observations, an X-ray
source is found at the location of the black hole. With an absolute astrometric 
precision of 0.6\arcsec\ (90\% confidence level), the {\em  Chandra} locations 
are consistent with the  positions reported at other wavelengths (see
also Corbel et al. 2005 for \1h). As the focus of this paper 
is the spectra of the two black holes in quiescence, we refer the reader to 
the studies of the X-ray jets presented in Corbel et al. (2002; 2005), Tomsick 
et al. (2003a), Kaaret et al. (2003) as well as the Tomsick et al. (2001) black 
hole study for details of the {\em Chandra} data reduction.

\section{Spectral properties of \xte\ and \1h in quiescence}

We extracted energy spectra in the 0.3--8 keV energy range for the black holes 
in all {\em Chandra} observations using CIAO tools, and we fitted these spectra 
using XSPEC.  We used a circular source extraction region with radii of 
1.4\arcsec and 4\arcsec, for \1h\ and \xte, respectively.  We extracted background 
spectra from annuli with inner radii of 9\arcsec and 6\arcsec and outer radii
of 19\arcsec and 16\arcsec\ for \1h\ and \xte, respectively.  These regions were 
centered on the black hole positions as given by {\it wavdetect}. 
The source aperture size is adpated to enclose most of
the source counts. Background region  are adapted to cover a
sufficiently large empty region. Due to the low 
numbers of source counts for \1h\ (52 counts or less), we used the W statistic for 
fitting (Wachter et al. 1979; Arnaud, in prep.) the un-binned spectra. This is 
adapted from the Cash statistic (Cash 1979) and is valid for background subtracted 
spectra.  For \xte, which is brighter than \1h, we re-binned the spectra in order 
to have enough counts in each bin to be able to use the $\chi^2$ statistic (with 
the exception of the {\em Chandra} observations on MJD 51777, 51798 and 52444, 
for which we used the W statistic due to the low number of counts). 

These spectra are adequately fitted with a power-law model including interstellar 
absorption. We used this model because this spectral shape is typically seen for 
black holes in quiescence (Garcia et al. 2001; Kong et al. 2002), but we emphasize 
that we can not statistically rule out other spectral shapes. However, bremsstrahlung 
models provide un-physical temperatures as already mentioned in Kong et al. (2002). 
For \1h, we fixed the equivalent hydrogen absorption column density, N$_\mathrm{H}$ 
to the constant value (2.3 $\times$ 10$^{22}$ cm$^{-2}$) measured in {\em Chandra} 
observations of \1h\ during its 2003 outburst (Miller et al. 2005). For \xte, we 
left N$_\mathrm{H}$ as a free parameter in our fit.  As the values obtained (Table 1) 
are consistent with results from previous X-ray outbursts (9 $\times$ 10$^{21}$ 
cm$^{-2}$; Tomsick et al. 2001), we also fixed the column density of \xte\ to 
this value.  
This column density is inconsistent with the value inferred from the 
interstellar absorption lines (S\'anchez-Fern\'andez et al. 1999), but agrees
with the hydrogen column density deduced from X-ray spectra of \xte\ or 
its associated X-ray jets (Tomsick et al. 2001, 2003a; Kaaret et al. 2003),
and also with the Galactic column density along the line of sight (Jain et al. 1999).
The difference when compared to the optical measurement of 
S\'anchez-Fern\'andez et al. (1999) may be explained in various ways.  One 
possibility is that the X-ray absorption column density may be intrinsically higher 
than the optically-inferred value due to additional local material covering the 
X-ray source. Another possibility is that the absorption column density deduced 
from the interstellar absorption lines are artificially lower (due to saturation
for example, Hynes et al. 2004). We believe that the latter is more likely as 
\xte\ and its X-ray jets (which are separated by up to 30\arcsec ) have similar 
column densities. In addition, it is interesting to point out the absence of 
variation of the hydrogen column density between outburst and quiescence.  

For \1h, the best-fit photon indices are 1.3$_{-1.7}^{+2.1}$, 1.6$_{-1.3}^{+1.0}$ 
and 2.2 $\pm$ 0.6 for the observations on MJD 53048, 53089 and 53091, 
respectively.  Refitting these three data sets simultaneously (allowing the normalization
to vary) leads to a photon index of 1.96 $\pm$ 0.46 (with 90\% confidence errors). 
As there are more frequent observations of \xte, its X-ray spectrum is better constrained.
The fit results of the individual observations are reported in Table~1.  To provide
the best constraint on the spectral parameters, we simultaneously fit all seven datasets
simultaneously (again allowing the normalization to vary). Leaving the hydrogen column 
density as a free parameter leads to N$_\mathrm{H}$ = (8.8 $\pm$ 1.0) $\times $ 
10$^{21}$ cm$^{-2}$, which is consistent with previous estimates. With N$_\mathrm{H}$
frozen to 9 $\times$ 10$^{21}$ cm$^{-2}$ (Tomsick et al. 2001), we obtain the best 
constraint on the photon index for \xte\ in quiescence, $\Gamma$ = 2.25 $\pm$ 0.08 
(with 90\%  confidence errors). Figure 2 shows the 68\%, 90\% and 99\% error contours 
allowing two parameters (N$_\mathrm{H}$ and $\Gamma$) to vary.  The results indicate
that these parameters are very well constrained by our observations. 

The photon indices in all {\em Chandra} observations of \xte\ (Figure 3) are consistent with 
the value obtained by fitting the spectrum from the observation with the brightest flux
(on MJD 52345) alone.  This indicates that the X-ray spectra of \xte\ are consistent with 
the same spectral shape for all the observations, even though the flux varies by a factor
of at least 16. 

The fact that the X-ray flux varies between the observations indicates
that accretion onto the black hole has not stopped in quiescence. We use the {\em Chandra} 
observations of \xte\ on MJD 52344 to construct a light-curve of \xte\ with a time 
resolution of 1 ks (Figure 4). A fit with a constant level leads to a $\chi ^2$ of 74 for 
27 degrees of freedom, clearly indicating that there is also significant X-ray variability on short 
($\sim$ ks) time-scale.   

The un-absorbed 0.5--10 keV fluxes for \1h\ and \xte\ are reported in Table 1. For 
\1h, we fixed the power-law photon index to a value of 2.0, whereas for \xte, the 
photon index was left at its fitted value. For both sources, the column densities
were fixed as indicated previously. The quoted errors are based on the numbers of 
source and background counts and Poisson statistics.

\begin{table*}[H!]
\begin{minipage}[t]{\textwidth}
\caption{{\em Chandra} observations of \xte\ and \1h : Best fitting spectral parameters for a power-law model.
All quoted uncertainties are 90\% confidence ($\Delta  \chi^2$ = 2.7 for one parameter or $\Delta  \chi^2$ = 4.6 
for two parameters).  }
\label{table:1}
\scriptsize
\renewcommand{\footnoterule}{}  
\begin{tabular}{c c c l c c c c }   
\hline\hline
Source	& Exposure & Number\footnote{Number of total count within the source region in the 0.3--8 keV energy band}
  & N$_H$ & Photon  &  $\chi^2_\nu$ / dof & Cash\footnote{XSPEC Monte-Carlo simulations of 10,000 spectra based on the fitted model. 
We list the fraction of these simulated spectra that has a Cash statistic value less than the original data 
(for good fit, this number should be around 0.50). }    & F$_{\mathrm{0.5-10~keV}}$\footnote{Un-absorbed X-ray flux in 
the 0.5--10 keV energy band.} \\
 (Date)	 & (s) &  of counts& (10$^{22}$ cm$^{-2}$)      & index                  &               &    M.C. prob. & (10$^{-14}$ ergs s$^{-1}$ cm$^{-2}$) \\
\hline\hline
\noalign{\smallskip}
\xte\   & &                            &                        &               &          &                   \\
\hline
\noalign{\smallskip}
MJD 51777.4 & 4985  &  71  & 0.9 (fixed)         & 2.6 $\pm$ 0.4 & --          & 0.57     &  26.6 $\pm$ 3.5 \\
MJD 51798.3 & 4572  & 111  & 0.9 (fixed)         & 2.3$_{-0.2}^{+0.5}$ & --          & 0.57     &  46.7 $\pm$ 4.7 \\
MJD 52344.8 & 26118 & 1206 & 0.86$_{-0.11}^{+0.19}$ & 2.20$_{-0.18}^{+0.25}$ & 32.8/30       & --     &      \\
            &       &	   & 0.9 (fixed)            & 2.25$_{-0.10}^{+0.10}$ & 32.8/31       & --     &  93.7 $\pm$ 2.8\\
MJD 52444.5 & 18025 & 58   & 0.9 (fixed)            & 2.8$_{-0.9}^{+0.5}$ & --          & 0.55     &   6.8 $\pm$ 1.0\\
MJD 52542.0 & 24442 & 223  & 0.90$_{-0.23}^{+0.53}$ & 2.2$_{-0.5}^{+0.7}$ & 16.1/19       & --     &          \\
            & &	    & 0.9 (fixed)            & 2.23$_{-0.24}^{+0.25}$ & 16.1/20       & --     &  17.6 $\pm$ 1.2\\
MJD 52667.3 & 23683 & 254  & 0.90$_{-0.22}^{+0.47}$ & 2.2$_{-0.4}^{+0.6}$ & 13.9/19       & --     &              \\
            & &	    & 0.9 (fixed)            & 2.19$_{-0.21}^{+0.23}$ & 13.9/20       & --     &  21.3 $\pm$ 1.3\\
MJD 52935.6 & 47835 & 145  & 0.69$_{-0.22}^{+0.47}$ & 2.2$_{-0.5}^{+0.7}$ & 17.3/19       & --     &              \\
            &       &	   & 0.9 (fixed)            & 2.5$_{-0.3}^{+0.4}$ & 18.5/20       & --     &   5.7 $\pm$ 0.5\\
\hline
\noalign{\smallskip}
Combined & &	   & 0.88 $_{-0.09}^{+0.12}$  & 2.25$_{-0.13}^{+0.09}$  & 137/167       & --   &        \\
            & &	   & 0.9 (fixed)              & 2.25 $\pm$ 0.08       & 137/168       & --   &        \\
\hline\hline
\noalign{\smallskip}
\1h\  & &	&  (10$^{22}$ cm$^{-2}$) &          &                     &      & (10$^{-15}$ ergs s$^{-1}$ cm$^{-2}$)  \\
\noalign{\smallskip}
\hline
\noalign{\smallskip}
MJD 53048.0 & 17796 & 6  & 2.3 (fixed)     & 1.3$_{-1.7}^{+2.1}$ & --  &  0.33  &         \\
            &       &    & 2.3 (fixed)     & 2.0 (fixed)            & --  &  0.48  & 12.2 $\pm$ 5.2 \\
MJD 53088.9 & 28363 & 16 & 2.3 (fixed)     & 1.6$_{-1.3}^{+1.0}$ & --  &  0.38  &                \\
            &       &    & 2.3 (fixed)     & 2.0 (fixed)            & --  &  0.52  & 19.5 $\pm$ 5.3 \\
MJD 53091.5 & 40037 & 52 & 2.3 (fixed)     & 2.2 $\pm$ 0.6 & --  &  0.64  &                \\
            &       &    & 2.3 (fixed)     & 2.0 (fixed)            & --  &  0.62  & 50.3 $\pm$ 7.3 \\
\hline
\noalign{\smallskip}
Combined  & &	  & 2.3 (fixed)   & 1.96 $\pm$ 0.46 & --  &  0.62  &                \\
\hline \hline
\end{tabular}
\end{minipage}
\end{table*}


\section{Discussion}

\subsection{X-ray luminosity in quiescence}

These {\em Chandra} observations  represent the detection of \xte\ and \1h\ 
at their faintest level of X-ray emission. The un-absorbed 0.5--10 keV 
luminosity of \xte\ is 1.9 $\times$ 10$^{32}$ erg s$^{-1}$ (for a distance
of 5.3 kpc, Orosz et al. 2002) and 9.3 $\times$ 10$^{31}$ erg s$^{-1}$ for \1h\ 
(at a distance of 8 kpc). For a ten solar mass black hole, these values 
correspond to luminosities of 1.5 $\times$ 10$^{-7}$ L$_\mathrm{Edd}$ (for \xte) 
and 7.2 $\times$ 10$^{-8}$ L$_\mathrm{Edd}$ (for \1h).  We note that the mass
of the compact object in \1h\ is unknown, as well as its distance. However, 
the kinematic study of the large scale jets of \1h, as well as its location 
toward the Galactic bulge, are consistent with a distance to \1h\ of 8 kpc
(Corbel et al. 2005).  

These quiescent luminosity levels are in the upper range compared to the
the faintest detected levels of other black holes (Garcia et al. 2001, 
Kong et al. 2002, Sutaria et al. 2002, Hameury et al. 2003, Tomsick et al. 
2003b). 
We note that there is still 
the possibility that these sources were not observed in their true quiescent 
levels as the {\em Chandra} observations were carried out between outbursts. 
However, we note that for \xte, the {\em Chandra} observations took place 
during three distinct quiescent phases (Figure 1) and a similar quiescent 
X-ray level is detected for two of these phases. In addition, optical 
observations (Jain et al. 2001) confirm that \xte\ returned to 
quiescence on a timescale of months.  For \1h, we note that the last two 
{\em Chandra} observations were taken 2.5 days apart. The flux in the 
observation on MJD 53089 is consistent with the flux observed during the 
first {\em Chandra} observation taken 40 days before, which indicates that 
this is probably the quiescent level of \1h.  Between the second and third 
{\em Chandra} observations, the flux more than doubled in less than three 
days. This may be related to continuous accretion in quiescence or 
possibly means that this could be the onset of the outburst that
was detected by {\em RXTE}/ASM about 90 days later (Swank 2004).
Similarly, RXTE/PCA detected the X-ray  activity of \gx\ during its 2002 
outburst at least 40 days before the initial ASM detection (Homan et al. 2005).
This probably indicates that the {\em Chandra} observations 
at the lowest level 
are likely representative of the true quiescent phase of \xte\ and \1h .

\subsection{X-ray spectra and photon index in quiescence}

In addition to allowing for detection of these two black holes in quiescence, 
for \xte\ these observations also provide the most precise X-ray spectra of a 
black hole at such low luminosities (down to $\sim$ 10$^{-7}$ L$_\mathrm{Edd}$). 
We concentrate on the power-law model, which can be representative of various 
emission mechanisms (see section 4.4). With a photon index of 2.25 $\pm$ 0.08, 
the X-ray spectra of \xte\ are significantly softer than during the standard hard 
state (photon index of the order of 1.5 to 1.7; Tomsick et al. 2001). This 
trend was marginally apparent in the decay of the 2000 outburst of \xte\ (Tomsick 
et al. 2001), but our {\em Chandra} observations of \xte\ clearly confirm this 
softening in quiescence.

In fact, this is very similar to several other black holes for which an X-ray 
spectrum has been measured in quiescence. For the rest of this paper, we will 
concentrate on those systems for which the photon index has been measured 
with relatively high precision. We have collected results from the literature which
are presented in Table 2 (90\% confidence level for the uncertainty on the 
photon indices) and in Figures 5 and 6. 
The X-ray spectrum of \xteonze\ ($\Gamma$ = 2.02 $\pm$ 0.16) in quiescence (McClintock et al. 
2003) is also softer than during its standard hard state ($\Gamma$ = 1.7; Frontera et al. 2003).
When \gx\ is detected at very low X-ray flux (Corbel et al. 2003; Gallo et al. 2003b; Chapuis et al. in prep.),
the X-ray spectra are again consistent with being soft ($\Gamma$ = 1.99 $\pm$ 0.15; Chapuis et al. in prep)
like \xte\ or \xteonze. Similar conclusions can be drawn for A~0620$-$00, which has a 
photon index of 2.26 $\pm$ 0.18 in quiescence (McClintock et al. 2003). 
However, the X-ray spectra of a subset of sources stay hard or harden significantly
in quiescence compared to the standard hard state: V404~Cyg ($\Gamma$ = 1.55 $\pm$ 0.07
with N$_\mathrm{H}$ frozen to optical value; Kong et al. 2002) and V4641~Sgr
($\Gamma$ = 0.2 $\pm$ 0.9; Tomsick et al. 2003b). With a photon index of
1.30$_{-0.41}^{+0.34}$ (Hameury et al. 2003), \gro\ may also be considered as
being hard in quiescence. 

It is interesting to note that the three BHs with the hardest X-ray spectra in quiescence
are those with the largest orbital period (Fig. 5). From our sample, all X-ray spectra 
from quiescent BHs with orbital period of less than two days are consistent with a soft 
power-law, whereas above this limit the power-law is hard. The weighted average photon index for 
the four ``soft'' sources is $\Gamma$ = 2.18 $\pm$ 0.06, whereas $\Gamma$ = 1.53 $\pm$ 0.07 
for the three ``hard'' sources. The difference between the two samples is significant at 
more than seven $\sigma$, clearly indicating that the photon index may depend on the 
orbital period. However, our sample only includes seven sources with the hard 
sample dominated by V404~Cyg, and it needs to be confirmed by further observations of BHs in 
quiescence. \grs, with an orbital period of 833 hours, would be a very good target if it 
returns to quiescence.

But as noted for \xte\ (section 3), the knowledge of the interstellar absorption column 
density is an important parameter for fitting the X-ray spectra. Indeed, lower column 
density leads to harder spectra (Fig. 5 in Kong et al. 2002). Some of the photon 
indices quoted above are deduced from X-ray spectra with a fixed column density.  We 
should now look if fixing the interstellar absorption could lead to an artificial bias 
in our sample.  For \xte\ (this work) and \gx\ (Chapuis et al. in prep.),  
the deduced N$_\mathrm{H}$ (as a free parameter) is 
consistent with previous measurements. For \xteonze\ and A~0620$-$00, McClintock et al. 
(2003) used a fixed column density; however, as N$_\mathrm{H}$ is already very low for
both of these sources, higher values would lead to even softer spectra. The 
X-ray and optical values of N$_\mathrm{H}$ for A~0620$-$00 and \xteonze\ are consistent 
with each other (Kong et al. 2002, McClintock et al. 2003).  For \gro\ (Hameury et al. 
2003) and V404~Cyg (Kong et al. 2002), X-ray fitting may indicate a slightly higher 
N$_\mathrm{H}$ when compared to the optical measurement.  With N$_\mathrm{H}$ free, the 
photon indices are softer with $\Gamma$ = 1.54$_{-0.7}^{+1.2}$ for \gro\ and 
$\Gamma$ = 1.81 $\pm$ 0.14 for V404~Cyg. For V4641~Sgr, there is no independent 
measurement of the X-ray column density due to the very fast transient nature of the 
source. To summarize, if we take into account the influence of the uncertain
interstellar absorption column densities on the determination of the photon indices, 
the difference in photon indices is significant at the three sigma level.

If this trend is confirmed, it is of interest to understand the possible difference 
between the short and long orbital period systems.  As outlined in Menou et al. (1999 
and references therein), one of the obvious differences is the mass transfer rate 
between the companion star and the accretion disk. Indeed, following Frank et al. 
(1992), for Roche-lobe overflow systems, the mass transfer rate may be driven by two 
separate mechanisms: loss of angular momentum through gravitational radiation and 
magnetic braking (j-driven systems) or expansion of the donor star as it evolves away 
from the main sequence (n-driven systems). The j-driven systems would be found 
preferentially in short orbital period systems. The bifurcation period between these 
two populations is expected in the range 0.5--2 days (Menou et al. 1999). Interestingly, 
Menou et al. (1999) have estimated the mass transfer rate for typical binary-evolution 
models (see their figure 3) driven by gravitational radiation and by secondary expansion.
If we compare the orbital period of the sources in our sample (4.1 hrs $<$ P$_\mathrm{orb}$ 
$<$ 155.3 hrs), we observe that systems with orbital periods in the range 4--30 hrs
would have similar mass transfer rates to within a factor of a few. In any case, the 
systems with the longest orbital period would have a much higher mass transfer rate,
and therefore, this could be an origin of the possible difference in photon indices.

Taking into account the mass of the black hole (Orosz 2003 and references therein),
we can further look (Fig. 6) to see if a correlation exists between the power-law photon 
indices and the quiescent X-ray luminosities (expressed in Eddington units).  The X-ray 
fluxes used in this graph are from Table 2 and represents the lowest level of X-ray emission
reported for these sources. This figure 
illustrates that the sources with the softer or harder spectra do not occur in a specific luminosity 
range. 

\begin{table*}[h!]
\begin{minipage}[t]{\textwidth}
\caption{Parameters of our sample of quiescent black holes: photon index and orbital parameters.}
\label{table:2}
\scriptsize
\renewcommand{\footnoterule}{}  
\begin{tabular}{l c c c c c c c }
\hline\hline
\noalign{\smallskip}
Source       & Photon\footnote{90\% confidence level. References in text. See also discussion on the influence of the hydrogen column density on the photon index.}   & \multicolumn {2}{c}{Quiescent\footnote{0.5--10 keV unabsorbed X-ray flux. Adapted 
from quoted reference with PIMMS from Heasarc. 1: McClintock et al. 2003; 2: This work; 3: Gallo et al. 2003b; 4: Kong et al. 2002; 5: Tomsick et al. 2003; 6: Campana et al. 2001.}} & Orbital & Distance\footnote{1: McClintock et al. 2001; 2: Gelino et al. 2001; 3: Orosz et al. 2002; 4: Hynes et al. 2004; 5: Hjellming \& Rupen 1995; 6: Orosz et al. 2001; 7: Shahbaz et al. 1994.} & Primary\footnote{Estimated 1 $\sigma$ range from
Orosz (2003) and references therein. } & Secondary$^c$ \\
	     &    Index           & X-ray flux & Luminosity & Period        &          & Mass         &   Mass    \\
\cline{3-4}
\noalign{\smallskip}
             &  ($\Gamma$)        & (erg s$^{-1}$ cm$^{-2}) $ & (L$_\mathrm{Edd}$) &   (hrs)     & (kpc)    & (M$_\odot$)  &  (M$_\odot$)   \\
\hline
\noalign{\smallskip}
\xteonze\    & 	2.02 $\pm$ 0.16   & 8.6 $\times$ 10$^{-15}$[1] &  3.9 $\times$ 10$^{-9}$ & 4.08   & 1.8 $\pm$ 0.6~[1]  & 6.48--7.19  & 0.23--0.32 \\
A~0620$-$00  &  2.26 $\pm$ 0.18   & 2.6 $\times$ 10$^{-14}$[1] &  3.1 $\times$ 10$^{-9}$ & 7.75   & 1.16 $\pm$ 0.11~[2] & 8.70--12.86 & 0.48--0.97 \\
\xte \footnote{Orbital parameters from Orosz et al. (2002) taking into account the rotational velocity of the star.}	                 &  2.25 $\pm$ 0.08   & 5.7 $\times$ 10$^{-14}$[2] &  1.4 $\times$ 10$^{-7}$ &   37.03 & 5.3 $\pm$ 2.3~[3] & 9.68--11.58 & 0.94--1.64 \\
\gx  \footnote{Orbital parameters from Hynes et al. (2003)}  
	     &  1.99 $\pm$ 0.15   & 3.4 $\times$ 10$^{-13}$[3] &  3.5 $\times$ 10$^{-6}$ &  42.00 &  $>$ 6~[4] &  5.8        &   0.52    \\
\gro         &  1.30$_{-0.41}^{+0.34}$& 2.8 $\times$ 10$^{-14}$[4] & 4.3 $\times$ 10$^{-8}$ &  62.92& 3.2 $\pm$ 0.2~[5] & 6.03--6.57 & 2.23--2.74 \\
V4641~Sgr    &  0.2 $\pm$ 0.9     & 1.0 $\times$ 10$^{-14}$[5] & 1.2 $\times$ 10$^{-7}$ &  67.61  & 9.6 $\pm$ 2.4~[6] &  6.82--7.42 & 2.85--3.34 \\
V404~Cyg     &  1.55 $\pm$ 0.05   & 8.9 $\times$ 10$^{-13}$[6] &  6.5 $\times$ 10$^{-7}$ & 155.28 &  3.0 $\pm$ 0.8~[7] & 10.06--13.38 & 0.53--0.83\\
\hline \hline
\end{tabular}
\end{minipage}
\end{table*}

\subsection{X-ray spectra at intermediate luminosities}

Most of the X-ray Novae have not been observed/detected in X-rays in their
quiescent state. However, we note that the softening of their X-ray spectra
has been seen at flux level well above quiescence.  In the decay phase 
of their recent outburst, \xt\ (Corbel et al 2004), as well as XTE~J1908$+$094
(Jonker et al. 2004), are first observed to harden significantly in the hard state
(we note that this is also the case for \gx\ at intermediate luminosity, possibly
related to Compton reflection, Nowak et al. 2002).
However, a clear softening ($\Gamma$ = 1.96 $\pm$ 0.09), like in \gx , is then
observed in \xt\ in the decay at a luminosity of 2 $\times$
10$^{34}$ erg s$^{-1}$ (Tomsick et al. 2004). No information is available on the
quiescent spectra of \xt\  or XTE~J1908$+$094, but it would be interesting to see
if they also soften (as \gx ) in quiescence.
Similarly, 4U~1543$-$47 gradually softened at intermediate luminosity
($\Gamma$ = 2.22 $\pm$ 0.12), whereas the hardest spectra in the hard state
had $\Gamma$ = 1.64 (Kalemci et al. 2005). {\em XMM-Newton} confirmed this
softening with a photon index of $\Gamma$ = 1.94 $\pm$ 0.04 at a luminosity
level of 4 $\times$ 10$^{34}$ erg s$^{-1}$ (10$^{-5}$ L$_\mathrm{Edd}$)
(La Palombara \& Mereghetti (2005).  4U~1543$-$47 was not in its true X-ray
quiescence phase (at the time of the {\em XMM-Newton} observation) which is
believed to occur below 3 $\times$ 10$^{31}$ erg s$^{-1}$
(Garcia et al. 2001). However, the optical/infra-red monitoring (Buxton \& Bailyn 2004)
indicates that 4U~1543$-$47 reached its quiescent optical level only ten days after the
{\em XMM-Newton} observations. It is unlikely that the X-ray flux could have decayed by three
orders of magnitude on timescale of ten days. This may possibly point out that the optical
emission settled down to its quiescent level much before the X-rays. Other examples of
similar softening at intermediate luminosity during the decay phase include
XTE~J1748$-$288 (Kotani et al. 2000) and GS~1124$-$68 (Ebisawa et al. 1994).

\subsection{On the X-ray emission of black holes in quiescence}

One of the main results from our study is that a significant fraction of BHs have an
X-ray spectrum that softens at lower luminosity. The emission of quiescent BHs is 
usually consistent with an extension of the hard state properties to lower luminosities 
(Corbel et al. 2000, 2003; Kong et al. 2000; Tomsick et al. 2004), and
hard state models have been applied to explain the emission in the quiescent state.

Among these models, McClintock et al. (2003) explained the power-law photon index of 
$\Gamma$ = 2.02 $\pm$ 0.16  of \xteonze\ (one of the best studied black holes in quiescence) 
with an advection dominated accretion flow (ADAF) located inside a truncated standard thin 
disk. However, a pure ADAF model was insufficient to reproduce the X-ray spectrum of 
\xteonze\ and they had to reduce the mass accretion rate close to the black hole. 
Specifically, they assumed that the mass accretion rate varies as a function of radius 
as $\dot{M}(r) \propto r^{p}$ and obtained adequate fits only for $p \ge 0.2$.
The fate of the ``missing mass" is not described in McClintock et al. (2003) but could
possibly takes the form of an outflow (e.g. Quataert \& Narayan 1999), which is then
like the ADIOS model (Blandford \& Begelman 1999). This is also quite similar to Yuan 
et al. (2004), who added a jet contribution to the standard ADAF model in order to fit 
the spectral energy distribution of \xteonze\ in the hard state. Within this context, 
an X-ray power-law photon index of $\sim$ 2 is predicted for BH in quiescence (Yuan \& 
Cui 2005).  However, at low luminosity the accretion flow is known to be convectively 
unstable (Igumenshchev \& Abramowicz 1999; Quataert \& Gruzinov 2000), and this 
convection-dominated accretion flow (CDAF) has a very different structure than the 
standard ADAF, as convection significantly reduces the mass accretion rate 
(equivalent to setting ADAF p parameter to 1 in eq. 1 of McClintock et al. 2003). 
In this CDAF framework, the X-ray spectra are expected to soften significantly at 
very low luminosity (Ball et al. 2001) as observed in our study. 
So, it is likely that ADAF models would be able to reproduce the 
soft and hard spectra described above, by tuning the amount 
of outflow or convection with the p parameter. To reproduce the harder
spectra of longer orbital period systems, such as V404~Cyg, larger
values of the $p$ parameter, corresponding to stronger outflow or
stronger convection, would be required.

Alternatively, in standard sphere$+$disk model, a softening of the X-ray 
spectra would also be expected due to a decrease of the coronal optical depth as the 
mass accretion rate decreases (see, for example, the discussion in Tomsick et al. 2004).
In the magnetic corona model, active regions above the disk are responsible for 
producing the hard X-ray emission and constitute the base of the outflow (Merloni \&
Fabian 2002). In that model, the X-ray spectrum is expected to soften at lower 
accretion rate as the accretion power is carried away into the jets rather that being
used to heat the electrons  (Merloni \& Fabian 2002). Such a model could easily explain 
the soft spectra of our sources in quiescence. However, in this model, the inner 
boundary of the accretion disk does not vary with mass accretion rate. The characteristic 
frequencies of QPOs and of broad-band timing noise are known to vary at state
transitions and in the hard state (Tomsick et al. 2004; Kalemci et al. 2005), with 
lower frequencies generally occurring at lower luminosities. This is easy to 
understand in models where the inner disk radius increases at low mass accretion 
rates, but is difficult to explain in the models where the inner disk radius is
constant.

Alternatively, a new possibility for the origin of X-ray emission has emerged in recent 
years.  Indeed, a strong correlation is observed between the radio and X-ray emission
of \gx\ and V404~Cyg in the hard state, and this correlation seems to extend down to the 
quiescent level for these two sources (Corbel et al. 2003; Gallo et al. 2003). The radio 
emission originates from a powerful self-absorbed compact jet (Corbel et al. 2000), and 
it also seems possible that this jet may contribute to the high energy emission (Markoff
et al. 2001, 2003) of BH in the standard hard state. Furthermore, radio observations
show that the compact jet (at least for V404~Cyg) is maintained in quiescence (Gallo
et al. 2004). Even if the debate is still open regarding the contribution of the compact 
jet (Markoff \& Nowak 2004, Homan et al. 2005, Markoff et al. 2005) at high energy, we can check if the
jet model could explain the softening of X-ray spectra for BH in quiescence. To explain 
the X-ray spectra in the hard state, the electron energy distribution should have an 
energy index of the order of 2.2--2.4 (Markoff et al. 2003). If the electrons that 
could be responsible for the X-ray emission of BH in quiescence are above the cooling 
break due to spectral ageing, then the energy index of the electron distribution (above 
the break) would be 3.2--3.4, therefore increasing (by 0.5) the photon index to 
$\Gamma \sim$ 2.1 as observed in a large part of the sources from our sample. In this
picture, the reason for the change would be that the particles are not sufficiently 
re-accelerated at low accretion rate, which is indeed quite reasonable. Alternatively, 
the compact jet contribution may vanish at low luminosity giving the possibility that
thermal emission from the nozzle dominates the X-ray band
and therefore gives a different spectral shape. The hard spectra ($\Gamma \sim$ 1.6) 
would easily be explained if the compact jet properties are similar to the standard 
hard state.  This is consistent with the fact that these kind of spectra are observed 
for the sources (V404~Cyg, \gro\ and V4641 Sgr) that have a large accretion disk, and 
which may be a necessary condition to sustain 
a powerful compact jet.  Similarly, it is of some interest to note that acceleration 
processes are also believed to be inefficient in the jets of the low luminosity 
(10$^{-9}$ L$_\mathrm{Edd}$) supermassive black hole, Sgr~A$^*$, located in the center 
of our Galaxy (Melia \& Falcke 2001).  

\section{Conclusions}

We have observed and detected \xte\ and \1h\ during their quiescent states at their 
faintest level of X-ray emission with a 0.5--10 keV luminosity of 2 $\times$
10$^{32}$ (d/5 kpc)$^2$ erg s$^{-1}$ for \xte\ and 9 $\times$ 10$^{31}$ (d/8 kpc)$^2$ erg s$^{-1}$ for \1h.
Such levels of X-ray emission are in the upper range compared with the levels observed  in other
black hole systems.  The {\em Chandra} observations 
also provide the best constraint on the power-law index of \xte\ with an index 
of 2.25 $\pm$ 0.08. We focused our analysis on the power-law model as this spectral 
shape is typically seen for black holes in quiescence and can represent an
approximation for most theoretical emission mechanisms, but we emphasize that we 
can not statistically rule out other spectral shapes.  All {\em Chandra} spectra 
of \xte\ are consistent with a soft power-law; therefore, indicating that the 
X-ray spectrum softens at lower luminosity.  We bring to light that all systems 
with short orbital period ($\sim$ $<$ 60 hours) are consistent with a softening 
of their X-ray spectra in quiescence.  However, the long orbital period systems 
may be consistent with a hardening of their X-ray spectra, but confirmation of
this trend is required.  A possible and realistic test would be to obtain {\em XMM-Newton} 
or {\em Chandra} observations of long orbital period systems, like V4641~Sgr or 
\gro\, with a long exposure during quiescence. In addition, \grs, with an orbital 
period of 833 hours, may be a very good target if it returns to quiescence.
Simultaneously, further observations of short orbital period systems (like \xt, 
or 4U~1543$-$47 at lower luminosity) should be performed in order 
to test the softening of their spectra. We found that various classes of models 
(ADAF corona + jet, CDAF, sphere$+$disk, magnetic corona or jet models) are able to reproduce the 
softening of the spectra in quiescence, but we note that most of them need the 
presence of powerful outflow or significant convection in order to reproduce these soft X-ray spectra. 
This may increase the likelihood that outflows are present in the most frequent phase 
(quiescence) of black hole binary's activity and have significant influence on the
physics of these systems and  neighbooring  environment  (Fender et al. 2003, 2005).

\acknowledgments

S.C. would like to thank H. Falcke, R. Fender, A. King, E. Kording, and 
S. Markoff for discussions on various aspects of BH emission models. 
{\em RXTE}/ASM results are provided by {\em RXTE}/ASM team at MIT.  
We thank the anonymous referee for constructive comments.
PK acknowledges partial support from NASA Chandra grants
GO4-5038 and GO4-5039 and from a University of Iowa Faculty Scholar Award. 
JAT acknowledges partial support from NASA Chandra grant GO3-4041X.

\newpage

\begin{figure} \plotone{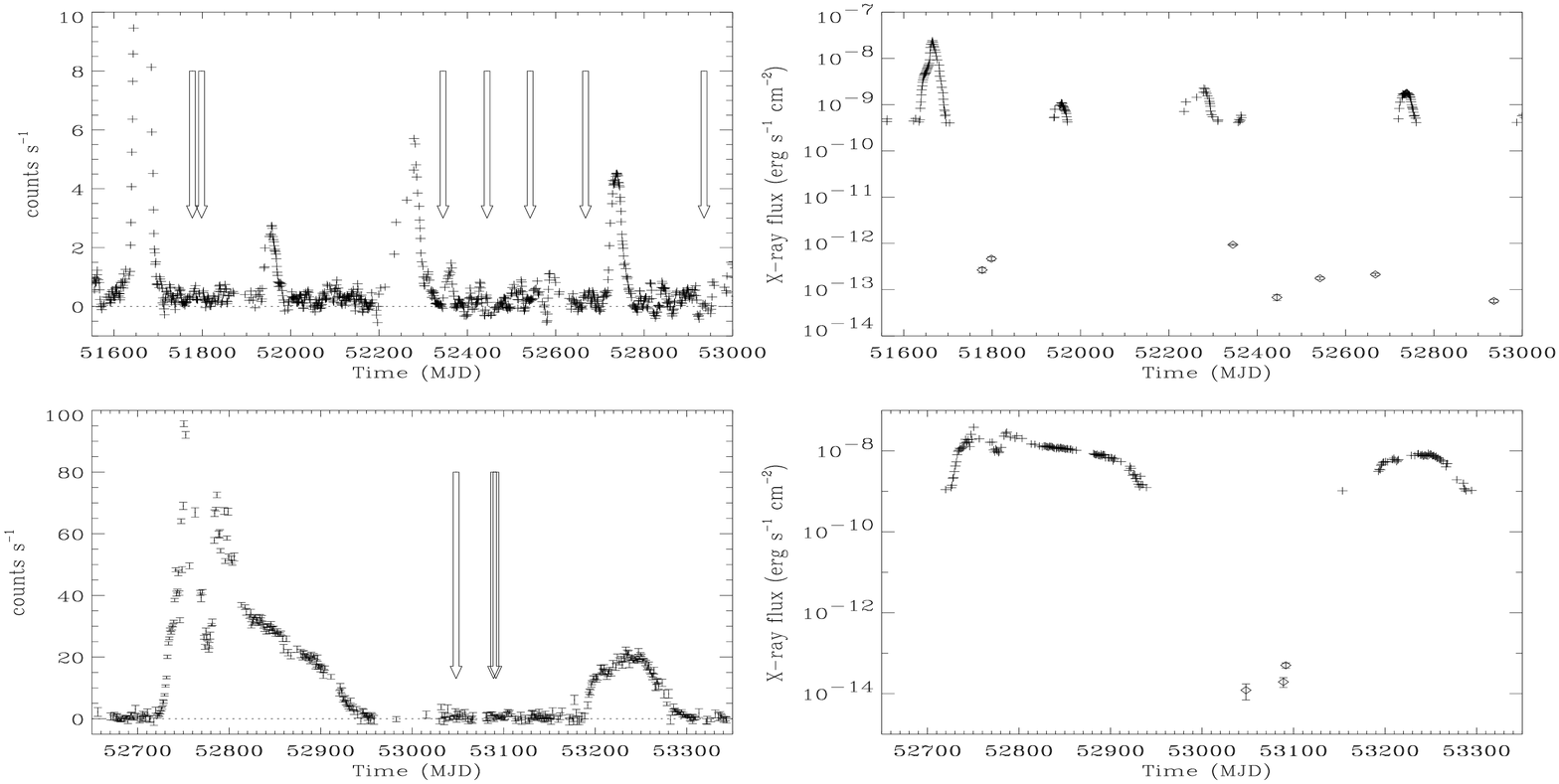}
\caption{{\em RXTE}/ASM 1.5-12 keV light-curve of \xte\ (Left/top) and \1h\ (Left/bottom)
covering the period of our {\em Chandra} observations. For the light-curve of \xte,
the points represent 5-day averages, highlighting the weaker outbursts observed in
2001, 2002 and 2003 (y-axis also truncated at 10 c/s for that purpose, the major
outburst in 1998-99 is not shown also), whereas we
plot daily averages for \1h. The arrows mark the time of the {\em Chandra} observations
which have all been conducted while the black holes were in (or close to) quiescence. In the
panels on the right, the ASM 1.5-12 keV lightcurves are expressed directly  (75~ASM~c.s$^{-1}$
=~1~Crab~=~3.0~$~\times$~10$^{-8}$~erg~s$^{-1}$~cm$^{-2}$) in flux units in order to
allow a direct comparison with the {\em Chandra} 0.5-10 keV un-absorbed flux measurements (diamond points).
The apparent ASM detection of \xte\ around MJD 52360 is likely an artifact due to the location of the
source close to the solar exclusion zone.
              }
\end{figure}

\begin{figure} \plotone{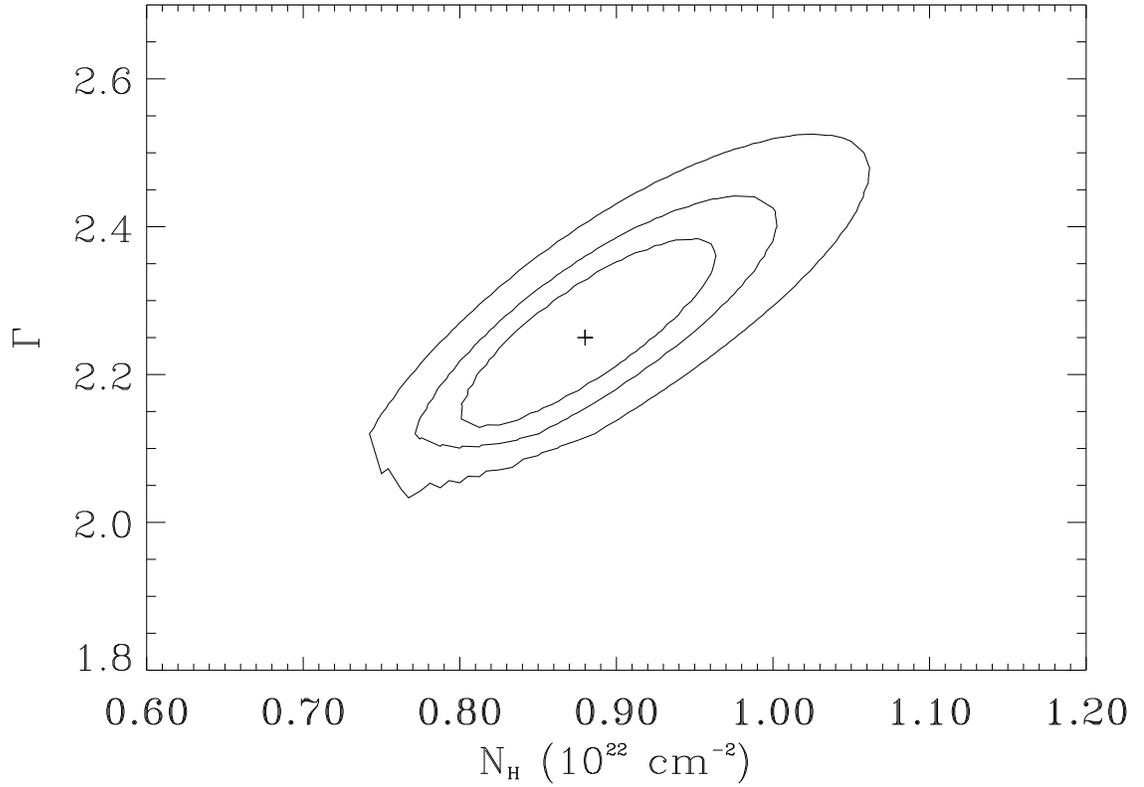}
\caption{Error contours for the hydrogen column density (N$_\mathrm{H}$) and
the power-law index ($\Gamma$) derived from the combined {\em Chandra} spectrum of
 \xte. The cross marks the location of the best-fit value, and 68\% ($ \Delta \chi^2$ = 2.30),
90\% ($ \Delta \chi^2$ = 4.61) and 99\% ($ \Delta \chi^2$ = 9.21) confidence contours are
shown.
        }
\end{figure}

\begin{figure} \plotone{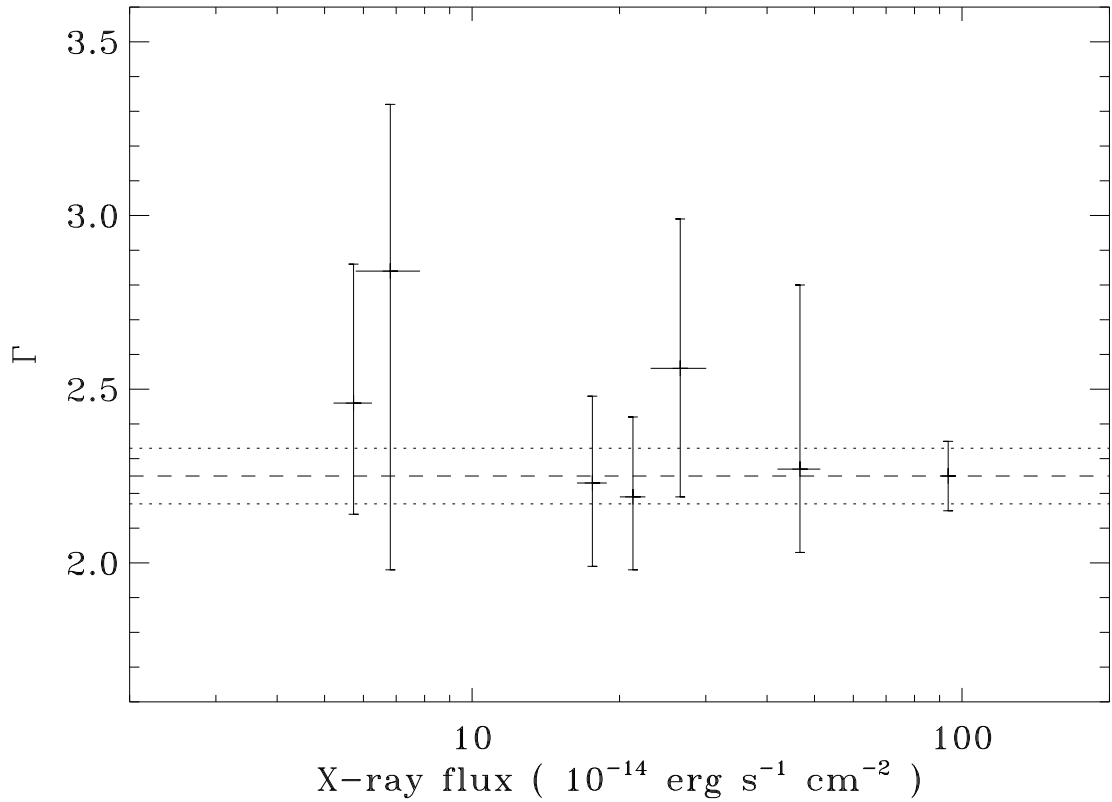}
      \caption{Evolution of the power-law photon index ($\Gamma$) versus the
un-absorbed 0.5--10 keV flux for all {\em Chandra} observations of \xte\ in quiescence.
The line indicates the best-fit value for $\Gamma$. All quoted errors are at the 90\%
confidence level.
        }
\end{figure}

\begin{figure} \plotone{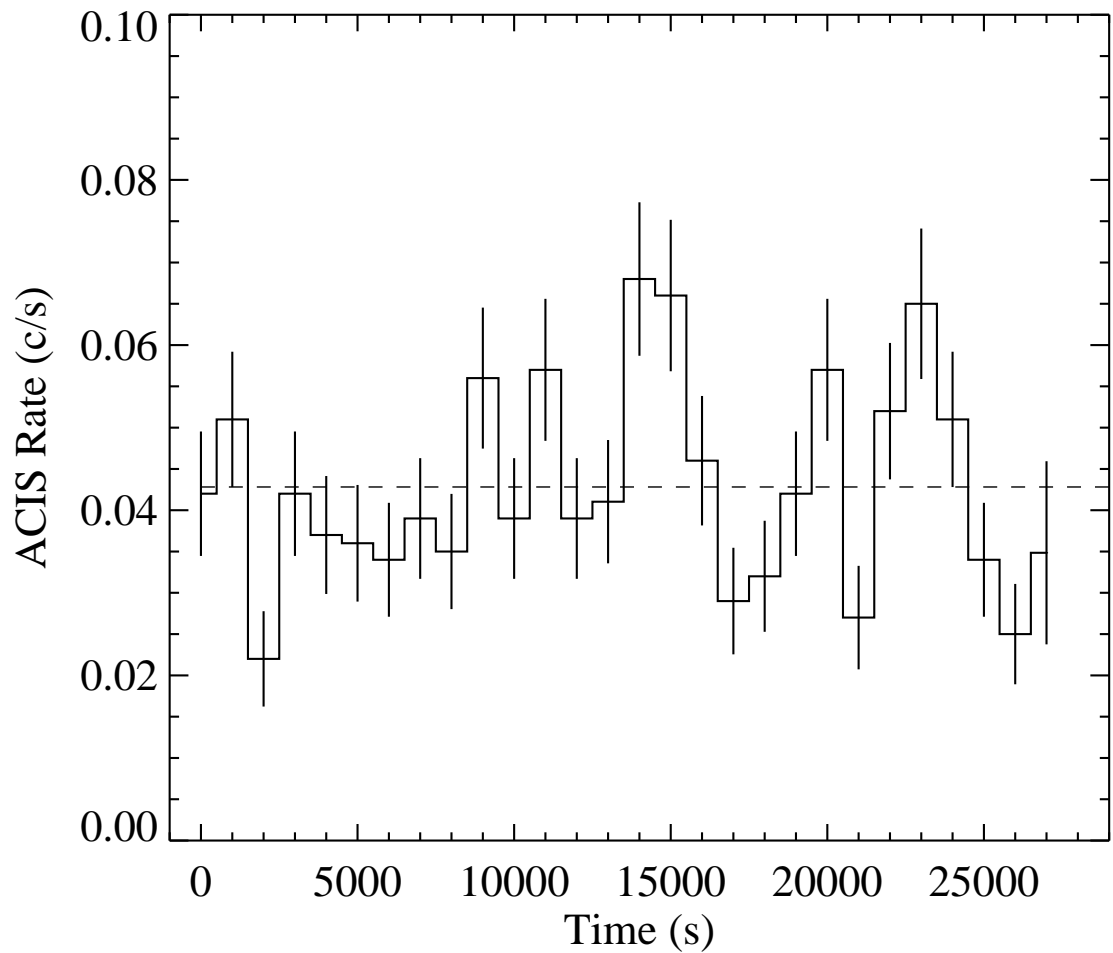}
      \caption{ {\em Chandra} ACIS-S light-curve of \xte\ on MJD 52344 in the 0.3--8 keV band with a
time resolution of 1 ks. The dashed line illustrates the fit with a constant level.
        }
  \end{figure}

\begin{figure} \plotone{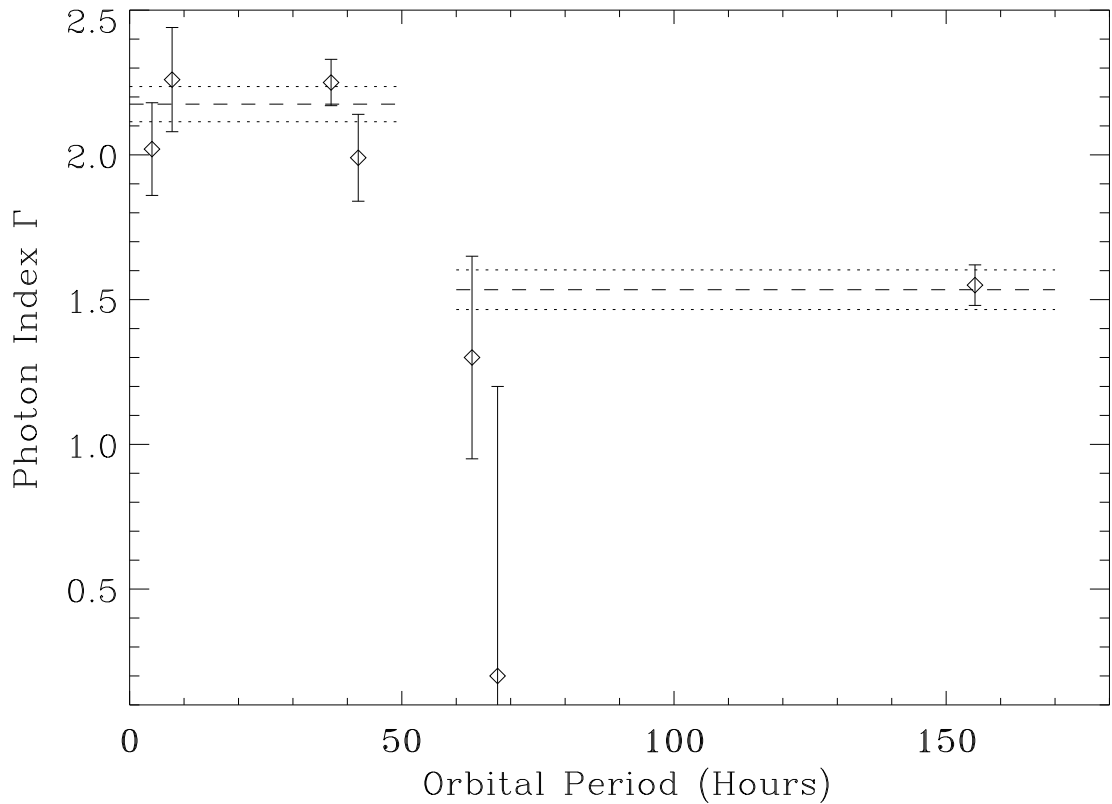}
      \caption{Evolution of the power-law photon index ($\Gamma$) versus orbital period
for \xteonze, A~0620$-$00, \xte, \gx, \gro, V4641~Sgr, and V404~Cyg (in order
of increasing orbital period) in their quiescent states. The lines (with associated
1$\sigma$ error) indicate the average value of the photon index for the two groups
(hard or soft -- see text) of sources.
        }
  \end{figure}

\begin{figure} \plotone{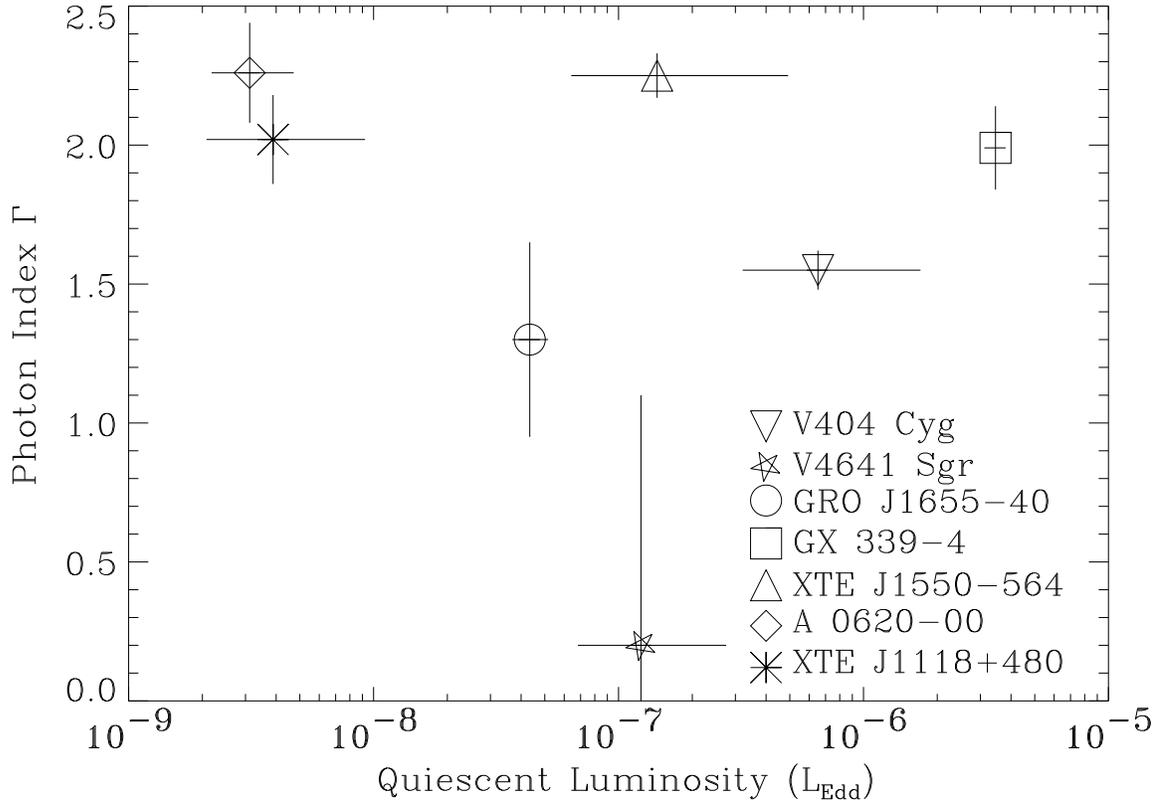}
      \caption{Evolution of the power-law photon index ($\Gamma$) versus the 0.5--10 keV
quiescent luminosity (in Eddington units). The uncertainty in the luminosity is based on
the uncertainty in the black hole mass and distance. A distance of 8 kpc associated
with a mass of 6 M$_\odot$ have been used for \gx\ (Hynes et al. 2003; 2004).
        }
  \end{figure}

\clearpage

\end{document}